\documentclass[12pt]{article}

\usepackage{scicite}
\usepackage{graphicx}

\usepackage{times}

\newcounter{lastnote}

\title{Measuring the phase velocity of twisted wavefronts}

\author
{Grace Richard$^{1}$, Holly Lay$^{1}$, Daniel Giovannini$^{1}$, Sandy Cochran$^{1}$, \\
Gabriel C. Spalding$^{2}$, and  Martin~P. J.~Lavery$^{1\ast}$
\\
\normalsize{$^{1}$School of Engineering, University of Glasgow, Glasgow, UK}\\
\normalsize{$^{2}$Department of Physics, Illinois Wesleyan, Bloomington, IL 61702, USA}\\ 
\\
\normalsize{$^\ast$To whom correspondence should be addressed; E-mail:  martin.lavery@glasgow.ac.uk.}
}

\date{}

\begin{document} 


\baselineskip24pt


\maketitle


\noindent \textbf{Traditionally one defines the speed of a wave as a property of the medium. Recent studies in photonics have challenged this idea, indicating that spatial shaping of the optical wavefront has can alter the arrival time of single photons when compared to an entangled reference photon. However, relying only on time of flight measurements leads to ambiguity in the measurement. Unlike photonics, one can directly measure the phase and intensity of sound, permitting unambiguous measurements. We developed a bespoke 28-element ultrasonic phased array transducer to generate short pulses that carrying orbital angular momentum. Through spatial mapping, we observed pulse modulation that indicates a localised change in phase velocity by over 10\,ms$^{-1}$. We present a geometrical argument for this effect, proposing that the phase velocity varies to compensate for the effective path length increase arising from spatial wavefront shaping. We expect that this pulse dispersion is a general effect for any spatially shaped wavefront and that it could be utilised to manipulate readings from pulse based sensor technologies such as SONAR, medical ultrasound, and underwater communication systems.}

\section*{Background}

At any given time, a propagating photon cannot be localised into a single well-defined position. Hence, the propagation velocity of a photon is a somewhat nuanced concept and the motion of the photon is more comfortably described in terms of the propagation of an extended wave-function with a particular group velocity. Changing the speed of light has captured researchers imaginations for several decades. Dispersive optical media are well known to reduced the group velocity of photons, with some astonishing experimental demonstrations that have slowed light down to less than 17\,ms$^{-1}$ \cite{Hau:1999fp,FrankeArnold:2011ec}.

In the last few years some interesting experiments in photonics have been conducted demonstrating that a similar effect is possible even in non-dispersive media, such as vacuum, via spatial shaping of a photon's wavefront \cite{Giovannini:2015kj,Bouchard:2016bb,Lyons:2018jk}. Component that such as a lens, Axicon or spiral phase-plate shape the wavefront of the beam that propagates through them by changing locally the Poynting vector \cite{Giovannini:2015kj,bornwolf}. Experimentally, the time of arrival of a single photon has been shown to change with respect to a comparative photon propagating over a straight optical path, as a result of spatial shaping, potentially indicating a change in the group velocity of the photon \cite{Giovannini:2015kj,Bareza:2016fk}. Similar suggestions of super- or sub-luminal behaviour have been documented in radio in close proximity to a radio antenna, where the speed tends back towards $c$ after long distance propagation \cite{Mugnai2000}.  Similarly in acoustics, waveguides and metamaterials have been demonstrated to be able to change the group velocity of a single acoustic pulse \cite{Zhu2016, Lemoult2013}. 

Many experimental measurement of super- or sub-luminal behaviour are based on temporal measurements. However, this leads to ambiguity in many cases, as it is experimentally challenging to distinguish between changes in relative optical path length and in the propagation velocity. Hence, acoustic waves offers a valuable parallel to investigate wavefront dynamics, as one can measure both phase and intensity simultaneously. Spatially-shaped wavefronts can be created through the use of controlled ultrasonic transducers arrays, and are used extensively for ultrasonic imaging \cite{Smith:1992cj}. Recent demonstrations have used controlled arrays to manipulate particles with acoustic tractor beams and to create touch-less haptic interfaces \cite{Demore:2014jv,Long:2014gs,Seah:2014gs}. Such arrays offer the ability to pattern the wavefront of sound dynamically in a manner similar to the way spatial light modulators do for light, and phased array antennae do for radio waves \cite{Konforti1998,Cochran:2006gs}. Specifically, recent experimental demonstrations have shown that large arrays of acoustic transducers can shape an acoustic wavefront to impart orbital angular momentum (OAM), as well as a wavefront that can be dynamically controlled \cite{Volke:2008gs,Skeldon:2008gs,Demore:2012gs,Brasselet:2012gs,Marzo:2018iy,Gibson:2018gs}.

\begin{figure}
  \centering
  \includegraphics[width=12cm]{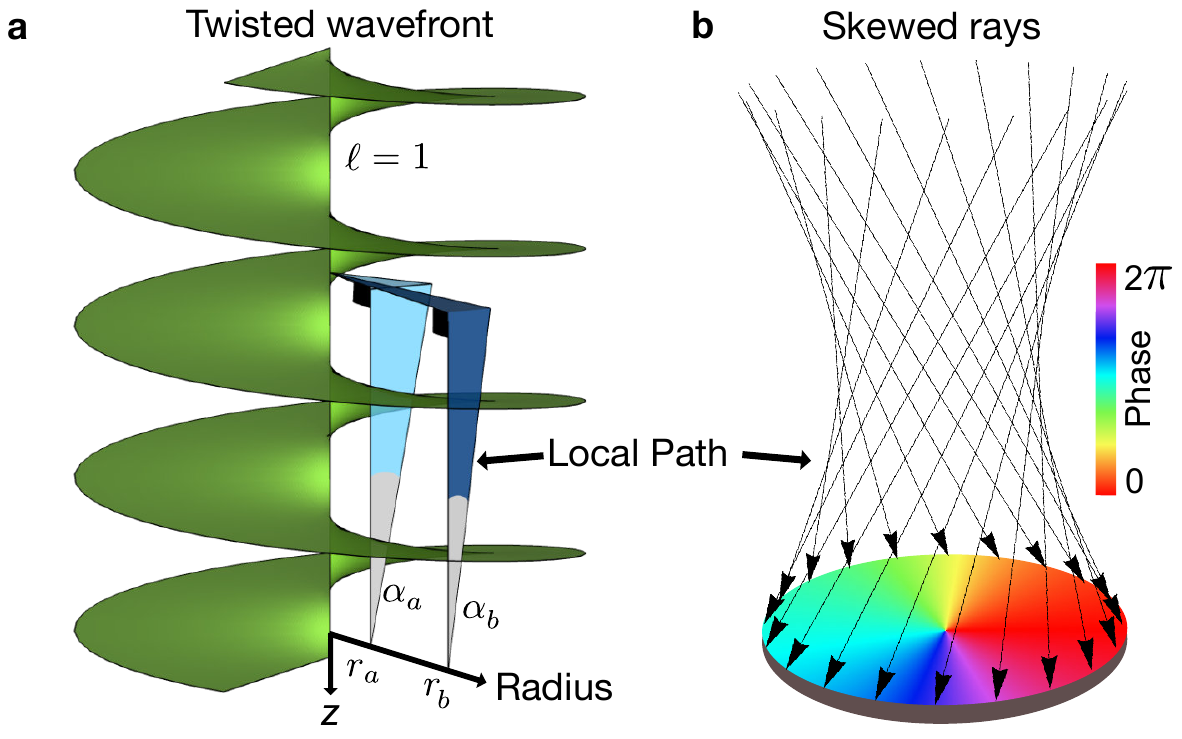}
\caption{A wavefront that carries OAM will spiral as it propagates through space.  \textbf{a} Here we show such a wavefront with a topological charge of $\ell =1$, which specifies both the number of intertwined helices and their pitch. These twisted wavefronts are associated with a change in the local ray direction that varies with radius. \textbf{b} Critically, these skewed rays have increased path length as compared to rays representing a planar wavefront.}
\label{skew}
\end{figure} 

In this study we will focus on wavefronts that carry orbital angular momentum (OAM). Such beams have had wide-reaching impact on the fields of optical manipulation \cite{Allen:2003aa,Andrews:2008aa,Rubinsztein:2017gs}, communications\cite{gibson:2004gs,willner:2012gs,lavery:2017gs,lavery:2018,hao:2015gs,Rusch:2018gs}, sensing \cite{lavery:2013gs,Arita:2013gs,Ritsch-Marte:2017gs,Moore:2018gs} and quantum information \cite{Leach:2002gs,Franke-Arnold:2004gs,malik:2014gs,Fickler:2016gs,Pan:2018gs}. Helical wavefronts were first noted to yield an OAM of $\ell \hbar$ per photon by Allen and colleagues in 1992 \cite{Allen:1992zz}, and are characterized by an azimuthal phase dependence of $\exp(i \ell \theta)$, where $\theta$ is the azimuthal coordinate and $\ell$ can be any integer. In 1999, Hefner and Marston demonstrated the presence of OAM in acoustic waves \cite{Hefner:1999ky}, whose twisted wavefronts result in a local change to the Poynting vector known as the skew angle $\alpha=\frac{\ell}{k \times r}$, where $k$ is the wavenumber and $r$ is the radial position \cite{Allen:1992zz}, see Fig. \ref{skew}. Our experimental findings indicate that this radial variation in the skew angle results in local change in the phase velocity of the structured wavefront.

\section*{Experimental Design}

We experimentally generated our structured wavefront with a bespoke array antenna with 28 independently controlled air-coupled ultrasonic transducers resonant frequency of 40\,kHz $\pm$ 1\,kHz (MCUSD14A40S09RS-30C, Premier Farnell, Leeds, UK). Each transducer was driven by a 0 - 5\,V square-wave digital signal with a frequency of 40\,kHz generated by a digital input/output device (USB-6255, National Instruments, Austin, TX, USA). The resonance of the transducer material leads to the generation of air coupled sine waves. The phase of the driving square wave can be individually tuned for each transducer. This tuning allows direct control of the local acoustic phase

\begin{figure}
  \centering
  \includegraphics[width=8cm]{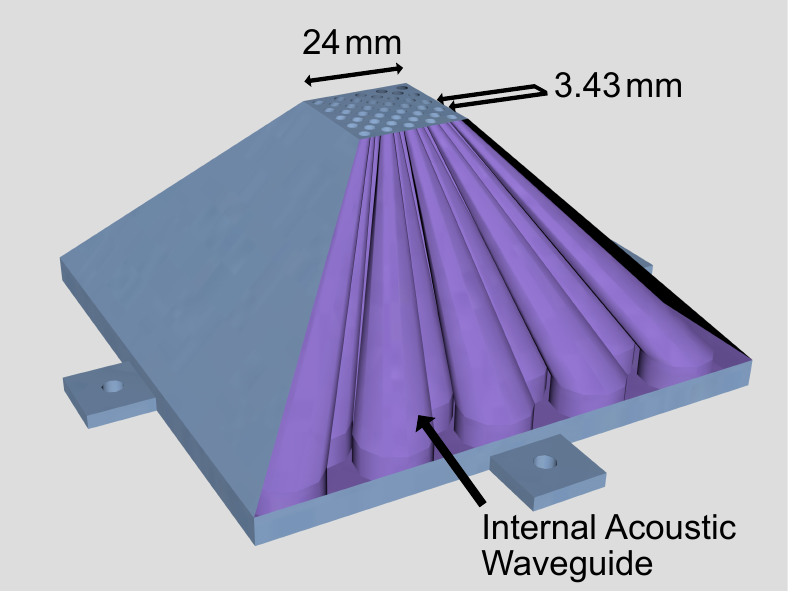}
\caption{ A 3D-printed waveguide array is used to reduce the effective separation between each of the acoustic sources. The 14 mm physical apertures of the transducers are reduced to a radius of 1.2\,mm with a center-to-center separations of 3.43\,mm through an array of tapered of air-core waveguides that redirect the acoustic energy.}
\label{array}
\end{figure}

The distance between adjacent acoustic sources is very important to generate a coherent acoustic wavefront as one can readily generate grating lobes that would represent significant artefacts in the generation of our desired wavefront. As our transducers have a physical aperture size of 14\,mm, we developed an integrated custom array of acoustic waveguides, manufactured through additive manufacturing and made from PLA (polylactic acid), to allow for close packing of the 28 acoustic sources, shown in Fig. \ref{array}. Each circular waveguide had an output diameter of  2.4\,mm arranged in a square grid with a pitch of 3.43\,mm. This spacing is half the acoustic wavelength, so grating lobes are suppressed. For the given drive signals, the peak sound pressure from each transducer is approximately 50\.dB, leading to a total sound pressure of 79\,dB. 

As phase-only modulation was used to generate our structured wavefronts, we required accurate calibration to phase match the waveguide outputs from the 28 independent transducers. Two sources of systematic phase error in our array antenna are the differences in path lengths of each of the acoustic waveguides and electrical impedance differences between each of the transducers. To overcome the path-length difference a specific time delay was added to each of the transducers based on precise measurement of the waveguide dimensions. To overcome impedance induced phase errors that resulted in sub-wavelength differences between each transducer, we directly measured the acoustic wave generated at each of the array output ports. The relative phase difference between each transducer was determined by computational comparison with a reference sine wave, matching the air-coupled waves generated by the transducers.  By adding the measured phase offset to each of the transducers, the array was fully calibrated to provide a flat acoustic wavefront.

\begin{figure}
  \centering
  \includegraphics[width=17cm]{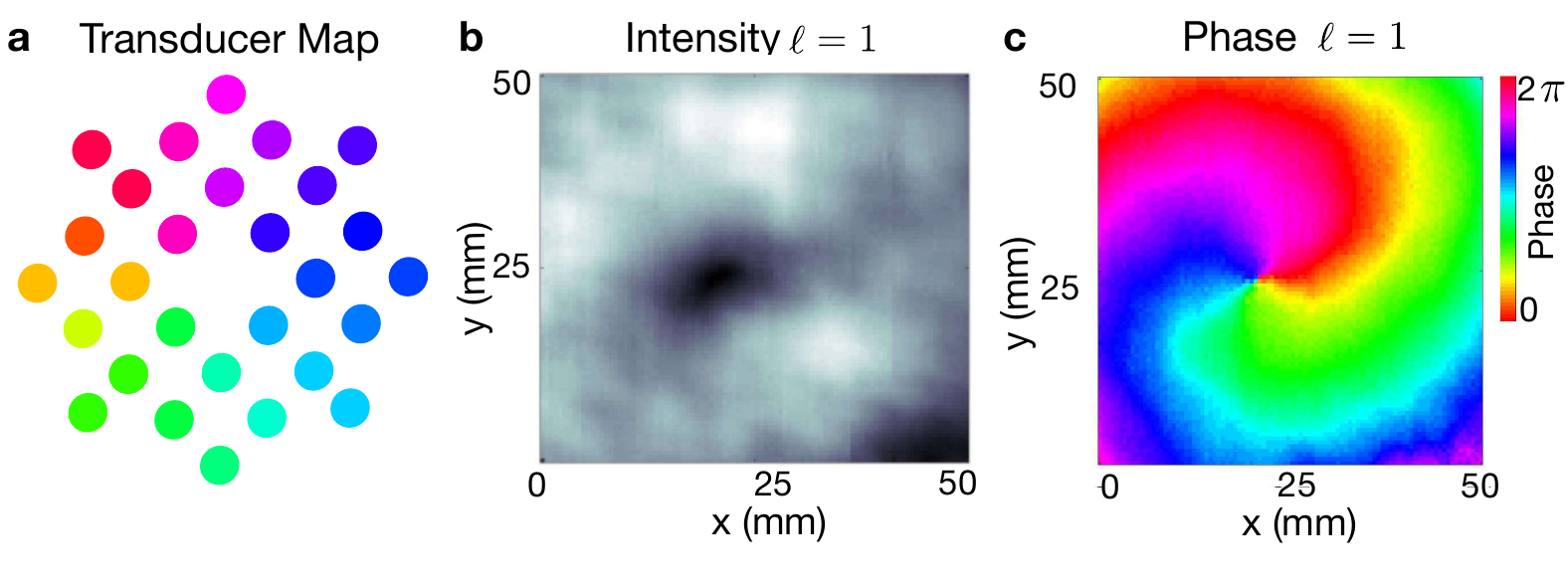}
\caption{\textbf{a} To generate a structured acoustic wavefront with $\ell = 1$, we independently control the phase of 28 acoustic channels  \textbf{b} After 60\,mm of propagation in air beyond the waveguide outputs, the measured intensity map has a clear node on along the propagation axis, which is a distinctive feature of beams that carry OAM. \textbf{c} At each spatial sampling point we also recover the phase of the acoustic signal, to map the phase profile of the acoustic wavefront. The resulting swirl is characteristic of a focused $\ell = 1$ beam.}
\label{setup}
\end{figure} 

To generate a specific acoustic wavefront, the phase of each transducer, $\phi$, is set based on the azimuthal position, $\theta$, such that $\phi = \exp (i \ell \theta )$, shown in  Fig. \ref{setup} \textbf{a}. Initially using a microphone with a 25-530\,kHz operational input frequency range (R3a 1232-1, Mistras, New Jersey, USA), the signal bandwidth for a range of $\ell$ values was determined to be within the range 38-42\,kHz. Subsequently, a high efficiency microphone with a suitable measurement sensitivity (MCUSD14A40S09RS-30C, Premier Farnell, Leeds, UK) was used record the presented data.  Using this microphone we measured the intensity and relative phase over a $50 \times 50$\,$mm^2$ region by scanning it with a spatial resolution of 0.5\,mm (Fig. \ref{setup} \textbf{b} and \textbf{c}). The microphone has a circular active area with a diameter of 14\,mm that was reduced to a diameter 1.2\,mm through the use of a PLA absorptive cone. 

\section*{Results and Discussion}

The propagation speed of a pulse can generally be separated into the speed of propagation of the pulses amplitude envelope i.e. the group velocity, $v_g$, and the propagation speed of the phase, i.e. the phase velocity $v_p$ \cite{Brillouin1960}. The phase velocity is determined with respect to the angular frequency, $\omega$, and the wave vector, $\textbf{k}$, as

\begin{equation} 
v_p=\frac{\omega}{\textbf{k}}.
\end{equation}

\noindent In the case of a mode with a plane wavefront, $\mathbf{k_{0}}$ is precisely defined as $| \mathbf{k_{0}} | = k = 2\pi /\lambda$. However, for structured modes there is a local variation in $\textbf{k}$ arising from the phase change across its wavefront \cite{Allen:1992zz}.  For beams that carry OAM the wavefront is locally skewed, leading to a variation in $\textbf{k}$ that is dependent on $\ell$ and $r$. Hence, this allows a local change in the phase velocity that will vary such that 

\begin{equation} 
v_p=\frac{\omega}{\mathbf{k_{0}} \cos(\frac{\ell}{k r})}.
\end{equation} 

\begin{figure}
  \centering
  \includegraphics[width=16 cm]{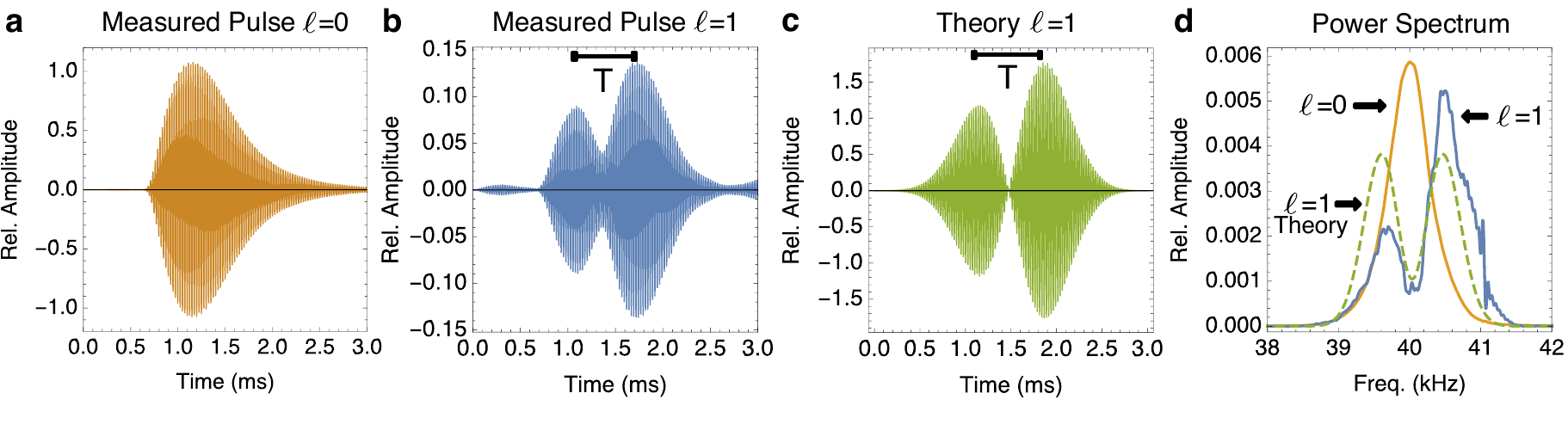}
\caption{\textbf{a} Measured pulse for an $\ell = 0 $ mode, where all 28 transducers are supplied with the same phase. \textbf{b} We measure a distinct amplitude profile when an an $\ell = 1 $ mode is produced by the acoustic array. \textbf{c} A expected pulse profile is calculated by considering our system properties and it is found the time T between the peaks in the modulated signal is close to that of the experimental signal.  \textbf{d} The observed beating pattern arises from the measured power spectrum shown. A continuous phase change is identically equivalent to a frequency shift, as the central wavelength is unchanged, our results indicate a change in phase velocity.}
\label{amplitude}
\end{figure}

To investigate the spatial change of $v_p$ across the shaped acoustic wavefront, we generated a single pulse with a length of approximately 1\,ms. A change in the phase velocity will result in modulation of the amplitude profile, $\psi$, of the propagating pulse at a particular position along its path of propagation, $z$. The amplitude profile of the pulse can be generalised to be

\begin{equation} 
\psi= 2 \cos(\Delta \mathbf{k} z - \Delta \omega t ) \cos(\mathbf{k_0} z-  \omega t)
\end{equation} 

\noindent where the change in phase velocity is  $\Delta  v_p=\Delta \omega /\Delta \mathbf{k}$ and $t$ is time. We can readily make an experimental acoustic measurement of this pulse amplitude profile as seen in Fig. \ref{amplitude}. For a flat wavefront, $\ell = 0$, we measured an elongated pulse with a centre frequency of 40\,kHz, Fig. \ref{amplitude} \textbf{a} and \textbf{d}. The pulse elongation arises from the impulse response of the transducers, that are designed to have a 40\,kHz resonant characteristic.  

For a shaped wavefront, with $\ell = 1$, we measured a pulse profile indicating modulation when our microphone is placed at the center of the twisted acoustic mode, Fig. \ref{amplitude} \textbf{b}. This change in amplitude indicates a change in the interference between composite frequency components comprising the pulse itself, and the superposition of wavefronts that combine at the microphone.  By considering the Equations 2,3 and the physical aperture size and position of our microphone, we computed an expected amplitude profile with similar shape, Fig. \ref{amplitude} \textbf{c}. In the frequency domain, it can be seen in Fig. \ref{amplitude} \textbf{d} that the center frequency of the modulated pulse is at the carrier frequency 40\,kHz. We expect the slight difference measured frequency components for $\ell = 1$  as seen in Fig. \ref{amplitude} \textbf{d} is from the spectral efficiency of our microphone and harmonics from background noise. It should be noted similar pulse modulation has been observed with the  25-530\,kHz microphone microphone, however with higher level of electrical noise due to lower microphone sensitivity.

\begin{figure}
\centering
\includegraphics[width=15cm]{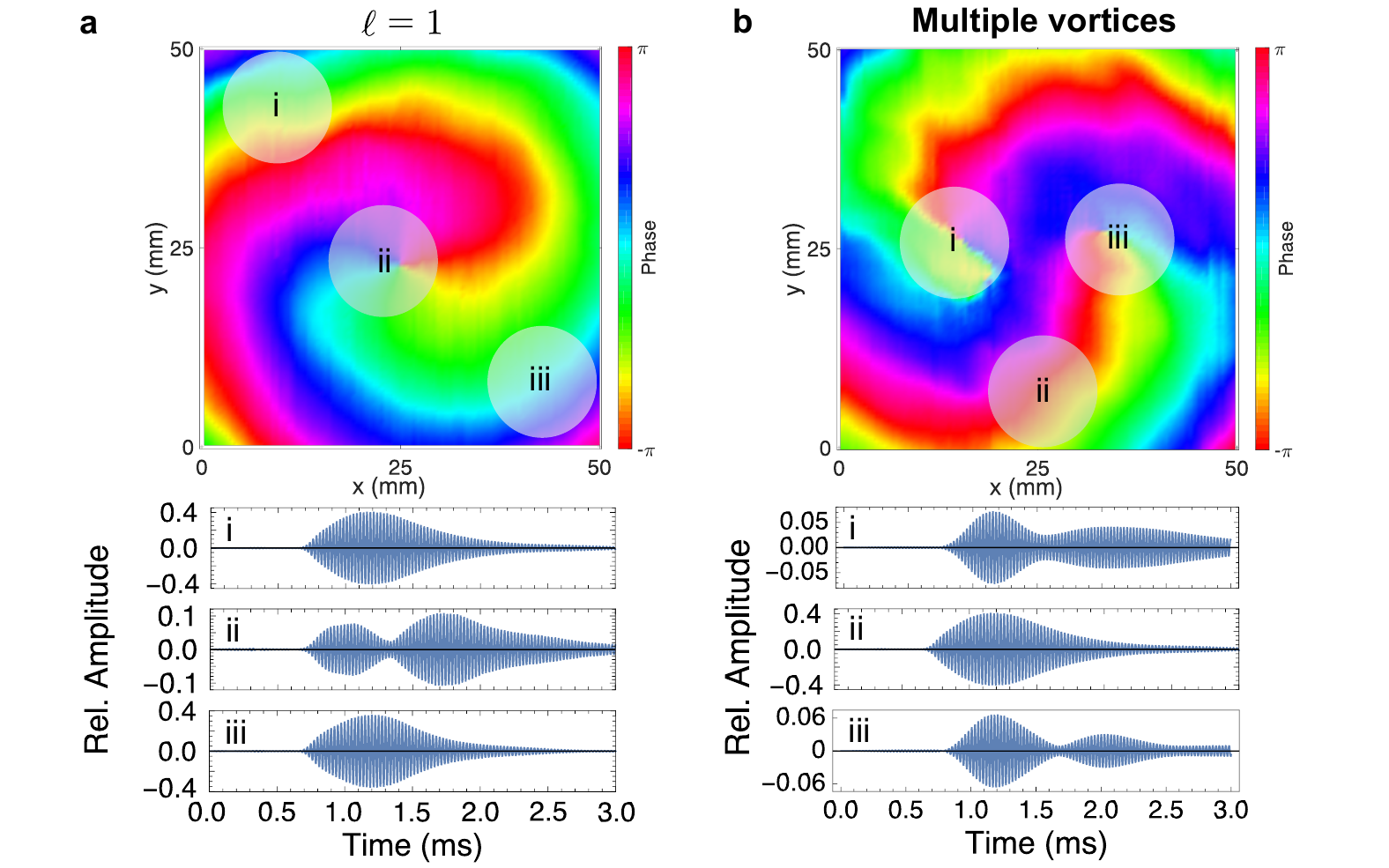}
\caption{The phase profile of the wavefront changes spatially across the aperture of the acoustic beam. Such a variation in local phase changes, locally, the direction of the wavefront and as a result changes the speed of sound locally. By relocating our microphone to several fixed positions we observe a distinctive change in the amplitude envelope at the centre of the mode for $\ell=1$, \textbf{a}, and two vortex nodes for a wavefront with multiple vortices, \textbf{b}.}
\label{oam2}
\end{figure}

From Equation 2, we expected the phase velocity to change with radius. Hence, we repositioned our microphone to record the full pulse profile at several fixed locations. It can be seen in Fig.\ref{oam2} \textbf{a} that the pulse amplitude appears similar to the $\ell = 0 $ case at larger radii. This is consistent with a smaller change in phase velocity at these positions in the wavefront. To confirm this observed amplitude modulation effect arises from spatial phase profile and not simply from a particular impedance response for a select few transducers placed near the middle of our array. We generated a more complex wavefront comprising two spatially separated acoustic vortices. This is achieved by programming our phase array antenna to generate an $\ell = 2$. However, as our system has relatively low spatial resolution  this leads to a spatial splitting of these vortices \cite{lavery:2018}. Moving the position of the microphone to measure the pulse profile at the centre of each vortex shows a pulse modulation similar to $\ell = 1 $,  Fig. \ref{oam2} \textbf{b}.

\begin{figure}
  \centering
  \includegraphics[width=15cm]{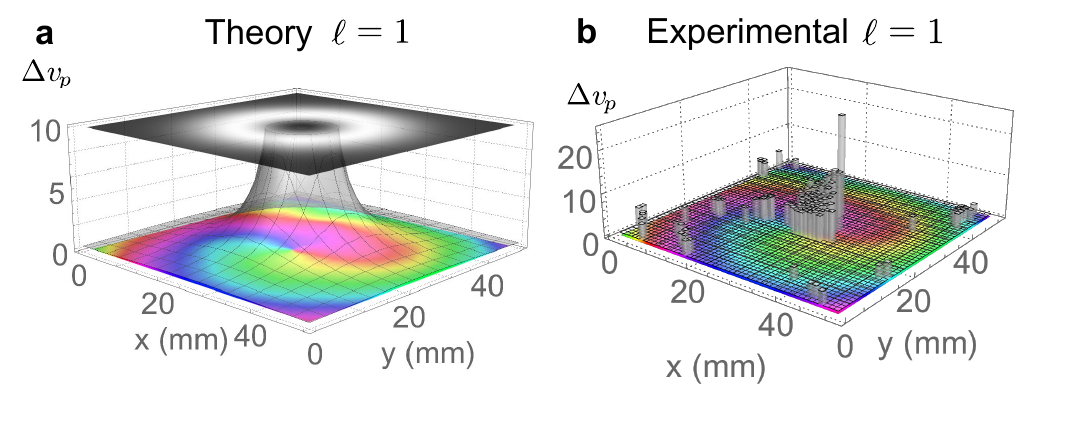}
\caption{\textbf{a} $v_p$ variance as determined using Equation 2. The intensity profile of the beam limits the maximum possible observable $\Delta v_p$. \textbf{b} Through measuring the amplitude profile of acoustic pulses in 1\,mm steps across a $50 \times 50$ \,mm$^2$ square aperture we can determine the local variation in the speed of sound. Both intentional structuring and phase aberrations contribute to the increase in the local $v_p$.}
\label{speed}
\end{figure} 

One can assess the $v_p$ difference across the measured field from the time delays between peaks in the pulse amplitude profile.  Two waves with slightly different frequencies will beat to an amplitude variation with $1/T = f_{r_a} - f_{r_b}$. This principle is central to the field of Doppler velocimetry, where the measurement of this beat is used to determine the speed of a moving object \cite{Durst1976}. Hence, through determination of the time between the pulse amplitude peaks, we calculate the speed change at any particular location in the acoustic wavefront. To calculate the local speed of sound, we measure $T$ at 1\,mm steps across the entire $50 \times 50$\,mm$^2$ sampling aperture, Fig.\ref{speed}. In certain cases only one peak is detected. However, the effective $T$ can be calculated through fitting of the theoretically expected peak position movement with respect to a pulse generated with a planar wavefront. By computing the expected phase velocity, we expect to see an increase in $\Delta v_p$ at the centre of the acoustic vortex \ref{speed}\textbf{a}. The intensity profile of the wavefront will limit the maximum observable phase velocity. Our results show a clear increase in the speed of sound close to the vortices for both $\ell =1$, see Fig. \textbf{b} respectively. However, we observe several locations at higher radius with increased phase velocity, which we attribute to local phase aberrations in the wavefront. We observer a peak $\Delta v_p = 25.0\,ms^{-1}$ and average $\Delta v_p=1.1\,ms^{-1}$ for $\ell=1$.

\section*{Conclusions}

Airborne acoustic fields have afforded us the ability to fully explore the details of structured wavefront propagation in a way that is more difficult in optics. This has demonstrated local changes in $v_p$ with average value of $\Delta v_p=1.1$\,ms$^{-1}$ that results from a spatial shaping of a propagation wavefront.  Our current results do not indicate any super- or sub-liminal effects as previously reported; however, we do note that any group velocity change could be unmeasurable given the propagation length of our system and the temporal resolution of our detection system.

As our experimental results indicate a change in $v_p$ will yield modulation of a propagating pulse, similar to dispersion. Hence, controlled spatial variation in phase or spatial phase aberrations could result in measurement errors in the arrival time of a particular pulse. Such an error could be critically important for SONAR, RADAR and LIDAR, especially where only a subsection of the entire wavefront is collected by a detector. Further, the pulse amplitude variation could result in errors for experimental systems that lock to a reference signal, such as heterodyne detectors or Hong-Ou-Mandel interference. Utilising this effect could provide novel routes to masking, or altering, the expected position from position sensing systems and to enhance security applications in acoustic communications.

\section*{Acknowledgements}
All authors would like to thank Matteo Clerici, Mark Dennis, Lucia Uranga and David Philips for useful discussions. Martin P.J. Lavery would like to acknowledge the Royal Academy of Engineering, EPSRC and the Scottish University Physics Alliance for their support. The work was support by the Royal Academy of Engineering, EPSRC awards EP/P510968/1 and EP/N032853/1.



\clearpage


\end{document}